# An In-Depth Analysis of Cyber Attacks in Secured Platforms

Ozoh Parick[1], John K Omoniyi[2], Ibitoye Bukola[3]

[1,3] Department of Information Systems, Osun State University, Nigeria
[2] Department of Industrial and Systems Engineering, North Carolina Agricultural and Technical University, USA

**Abstract**

There is an increase in global malware threats. To address this, an encryption-type ransomware has been introduced on the Android operating system. The challenges associated with malicious threats in phone use have become a pressing issue in mobile communication, disrupting user experiences and posing significant privacy threats. This study surveys commonly used machine learning techniques for detecting malicious threats in phones and examines their performance. The majority of past research focuses on customer feedback and reviews, with concerns that people might create false reviews to promote or devalue products and services for personal gain. Hence, the development of techniques for detecting malicious threats using machine learning has been a key focus. This paper presents a comprehensive comparative study of current research on the issue of malicious threats and methods for tackling these challenges. Nevertheless, a huge amount of information is required by these methods, presenting a challenge for developing robust, specialized automated anti-malware systems. This research describes the Android Applications dataset, and the accuracy of the techniques is measured using the accuracy levels of the metrics employed in this study.

## 1. Introduction

### 1.1. Cyber Attacks

Cyber attack is a risk to businesses and organisations in the medical, financial sector, and the entertainment industry. The use of the Internet of Things in contemporary classrooms has increased, transforming traditional systems into an interactive experience. [1] explore cybersecurity implications of IoT usage in systems, to identify malicious threats and assess measures taken to protect sensitive data. The methodology consists of a qualitative and exploratory approach. Data was gathered using case studies, interviews, journals, and reports. The findings present gaps in cybersecurity awareness, inadequate facilities, and the nonexistence of IoT safety protocols. The study concludes with a proactive approach, improved procurement policies, effective risk assessments, and enhanced training functions.

Unique solutions must be introduced to tackle the growing threats in cyber for communication networks. [2] present the effectiveness of the MARL algorithm. The methodology used is the Codalabs algorithm, which automatically measures the agent in the communication environment. This study provides a relative threshold for future systems. Initial studies show that machine learning displays high value in its use.

This study is used for classifying different Grey hole attacks using a unique dataset, FAN GHETS24 [3]. The dataset is obtained from a series of connections within the network, produced via several simulations of FANETs. A limitation is that a large quantity of the data is redundant; therefore, it can be helpful to filter the data before classification. [4] propose an audience-based procedure to improve cybersecurity in discrete time. The technique presents methods for integrating results used in generating corrective actions, thereby enhancing computational efficiency in solving problems. The method used simulates a system for control inputs, which is dependent on a priori-based boundaries for error detection, validated using numerical simulations for real-world situations. The limitation of this study is that it does not harbor nonlinear outputs.

### 1.2. Alleviating Cyber Attacks

[5] present information on present challenges and a guide to cybersecurity systems within IoT frameworks. The study collects and analyzes important information by identifying primary sources. A study of the primary security mechanisms for the research is compared to the present study, highlighting vulnerabilities across various architectures, particularly in the context of ransomware, malware, and other cyberattacks. The study proposes potential future research enhancing cybersecurity in this evolving field. [6] considers the various challenges of securing digital systems, particularly the need for effective IoT devices. This paper offers insights into protecting digital systems. As threats increase, this paper provides defensive capabilities for digital systems applied to various environments [7]. These solutions compete with resource constraints, in addition to reduced computational power and memory. This paper presents a method for data reduction in choosing a symbolic data set by using the Monte Carlo method to demonstrate a comparable classification for testing IoT systems.

[8] assess information of security platforms for the public service using assessments free from interference. The study presents a secure end-to-end communication system to assess recommendations to customers. The study displays gaps and protocols to indicate improvements in the system. This work is a guide to enhance security infrastructure by managing vulnerabilities for a robust cybersecurity system. [9] provide a review of past studies on solutions distinct to different challenges in communication systems. The study evaluates various emerging technologies for these networks, issues facing the implementation of the systems, and solutions developed to counteract these challenges. The challenge of this study is overcoming regulatory challenges to achieve its potential.

### 1.3. Challenges of Methods

A study analyzing enhanced classification algorithms to investigate the finding, sorting, and storing of fragile data across security platforms [10]. The study demonstrates significant accuracy. This signifies the dependability of the technique in detecting practicable security concerns for different datasets. The industry's encryption systems are analyzed and continuously monitored to identify and mitigate vulnerabilities, thereby enhancing detection. The integration of important classification techniques with optimal practices proposes a reliable technique. [10]

present more efficient data security methods. This study authenticates several schemes across different categories using a systematic framework that focuses on their practicality, effectiveness, and security [11]. The study showed a diverse performance, needing normalized results. This study uses a systematic review to evaluate knowledge-based security measures for the system. In considering the limits of the proposed system, cross-modal and context-aware methods enhance the security of the system's network.

[12] present an enhanced specification-based method for incorporating a security-based method to monitor the behavior of the nodes in the network. This includes a packet-forwarding trust together with the specification-based method. The method utilizes security-based tools to ensure reliability. The results on the Cooja simulator indicate that the method achieves improved accuracy. [13] Investigate the exposition of classification, detection, and a comprehensive analysis of different data collection methods. The diversity of the study was increased when variables were combined, indicating that combined data is more effective. The study proposes that the digitization of records will provide vital information on biodiversity, and a combination of both will allow changes to be tracked, helping to monitor responses to environmental change. The study considers biases unique to each data source, species detection, diversity patterns, and biases for separate areas. [14] present a detailed evaluation of cybersecurity challenges and provide solutions to tackle these challenges. With growing security concerns, protecting fragile details is crucial to protecting privacy and trust. To address these challenges, a secure platform consisting of encryption methods is required to address cybersecurity challenges for the safety of systems across domains, hence the need to focus on ensuring the safety and integrity of security systems to reduce the effects of cyber threats on society. The gaps in this study involve prioritizing a detailed analysis for investigation.

The following are the contributions of this study.

1 An exploration of past literature
2 Collection of data for Android and Linux encoding
3 Collection of system call data

## 2. Background

This section gives an exposition of Android's architecture and an overview of its security history, accompanied by a background of cyber threats. This section introduces Finite State Machines (FSMs) with a formal definition and expands on FSMs by describing the challenges involved in supervisor reduction algorithms [15]. A description of topics relevant to this study is given in the following sections.

## 2.1. The Android Platform

This is a platform comprising an interface and various applications. The applications of Android include Email, SMS software, calendar, maps, web, contacts, and other features. The software uses Java and gives details of its platform architecture. Android provides an architecture for different component layers to work together.

Figure 1 displays the architecture for Android.

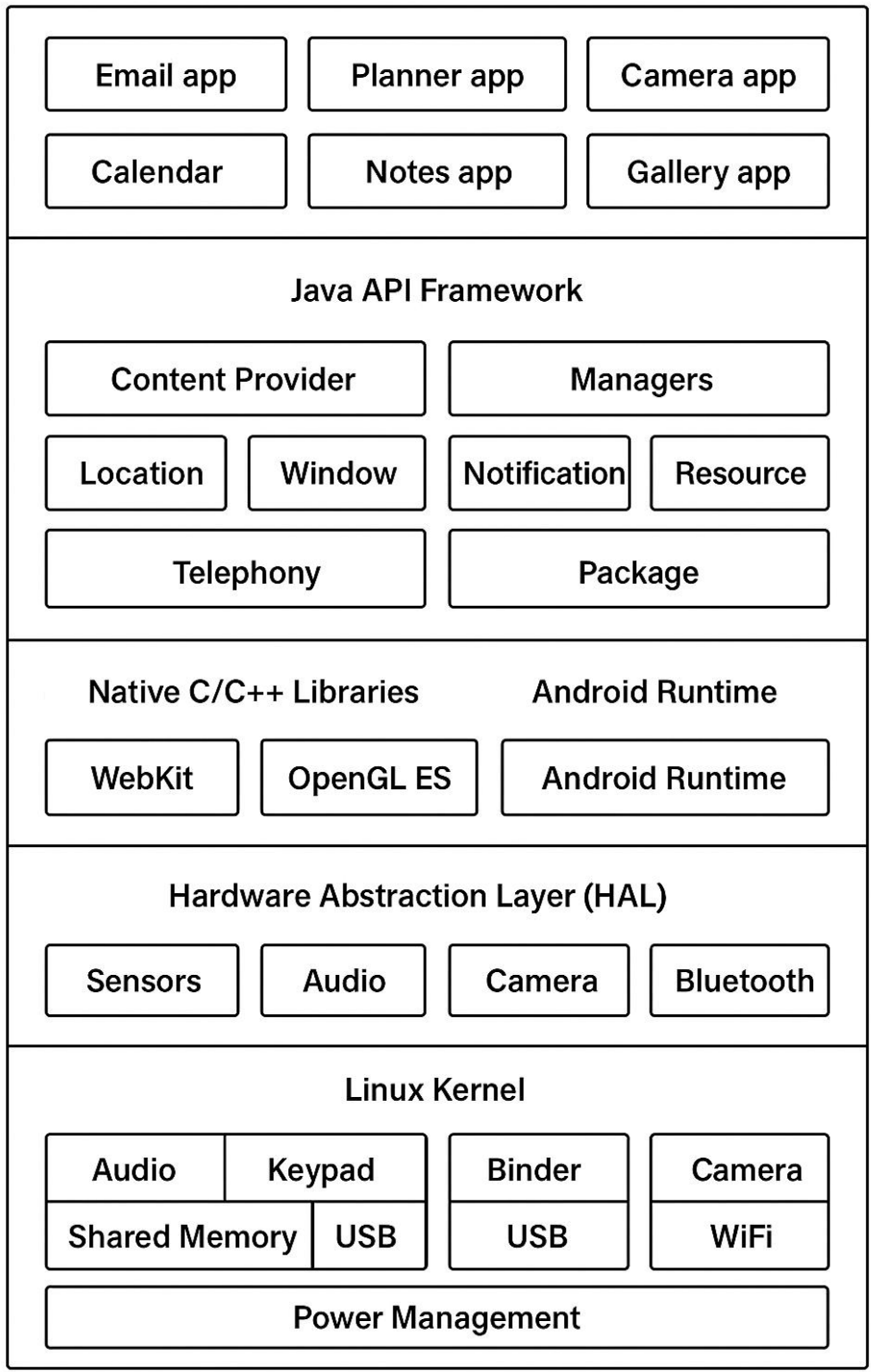

**Figure 1.** Android Software Architecture

The System Apps is the topmost layer, where applications are installed and synchronized with one another. The next layer is the Java API layer, which gives access to the Android Operating System written in Java, and displays primary components for developing the system. The third layer comprises primary C/C++ Libraries and the Android Runtime(ART). This enables entry to the system components. The second section is Android Runtime [16]. It presents an integrated approach that combines four techniques to improve detection accuracy, reducing and alleviating hidden security threats. This provides improvements in the predecessor, like enhanced error collection and diverse debugging features for the system. The security context comprises security threat identifiers that define permissions and access associated with the various processes. The next layer comprises the camera and sensors. For each interface, the library modules are packed when the Java API framework accesses hardware components. The end layer is the Linux kernel, which is the primary structure for Android that components depend on. The system is assigned a distinct secured code within the framework. This security code comprises security identifiers that label the permissions and access related to the individual process.

### 2.2. Chronicle of Android Mobile Protection

Since the advent of Android, there have been improvements and enhancements to its security. A classification describing the main techniques used to secure Android applications was presented by [17]. The paper analyzes software tools designed to assist developers in protecting their systems, utilizing a dataset consisting of malicious software. The study indicates that, despite the use of software protection techniques over the years, software is still utilized by a small portion of applications in the Android system. [18] present a method to enhance malware detection, thereby posing a protection against emerging threats. The findings indicate a positive development of Android security systems in the call network environment. The limitation of this study is the introduction of a ransomware attack mechanism that misclassifies generated examples.

This study addresses crucial requirements for a more effective cyber threat protection [19]. A unique defense method involving iterative mechanisms aims to evaluate the high sensitivity of perturbations. This technique detects cyber threats by measuring an increase in the loss function, separating them from regular samples. The study shows that the proposed method achieves a defense accuracy of 88.5% 90.7%, compared to the previous method. [20] Propose an active learning method that evaluates call monitoring to identify malware. The study shows hybrid model is more accurate compared to traditional methods (92.36 % and 85.9 %) and loss functions (22.5 % and 33.2 %).

### 2.3. Cyber Threats

The meaning of cyber threats is huge and evolving, as threat actors require methods to collect ransoms. This study limits the definition of cyber threat to malware, which holds the system or data to ransom [21], and discusses features and technologies exploring improvements in the system. Data vulnerability, malicious assaults, insecure systems, and security challenges were identified as cyber threats. The study identified improvements in the system as enhanced security and privacy in the digital cyberspace.

The growth experienced in mobile computing presents various forms of flexibility to different users. As a result, mobile users are vulnerable to cyber threats, and cybercrimes are evolving. This study focuses on vulnerabilities in the mobile computing industry, particularly tablets and smartphones [22], and provides responses that can be used to protect against such threats. [23] Review relevant literature to investigate the relationship between resources and malicious attacks and propose a categorization of the impact of social factors on mobile phones. The study facilitates a thorough understanding and provides comprehensive protection for the system, helping to develop unique solutions that protect users' privacy against various attacks.

Protecting data has become a valuable issue, as personal data has increasingly been targeted by cyberattacks. [24] Employs a comparative analysis to evaluate users' data in this era. This study shows enforceable guarantees and proposes compliance-based and post-control strategies that prevent security leaks and address their root causes. The study utilizes preventive mechanisms as tools to defend the system. [25] employ machine learning in monitoring threats for mobile networks. The study investigates network traffic and uses machine learning to determine malware patterns. The model is used to predict and stop threats in real-time, and these techniques can be used to prevent security breaches and control the mobile network. This method enhances the security of mobile networks by monitoring for potential security threats, reducing the time and resources required for incident response.

### 2.4. Introduction to Communication Networks

This application is effective for managing and enhancing a system[26]. It investigates traffic prediction, load balancing, and malware detection. Artificial intelligence can efficiently manage bandwidth in communication networks, protecting devices and enhancing security through real-time malware detection. The study addresses current and future challenges, including their impact on a secure and efficient network.

[27] Discuss the use of maximizing and improving cybersecurity. As communication networks change from 5G to 6G, the use of artificial intelligence in interpreting complex network issues becomes vital. This study assesses machine learning as a tool for enhanced performance and security. [28] present a systematic protocol that can spatially resolve data and identify altered datasets. This protocol typically takes around 20 minutes, and no specialized prior training are used to evaluate research conducted by sociological centres concerning social networks and user preferences, their age characteristics, and data from periodicals. As a result, the study compared the popularity rating of social networks. 6G will boost the relationship between human and digital domains, as a result, requiring the expression of primary performance indicators. The aim of [29] is to enhance current research on 6G computing, based on sustainable development goals. [30] present the characteristics and compile the number of active users. The variables are analyzed and evaluated.

### 2.5. Finite State Machines for Malware Detection

The ability to detect malicious threats in mobile phones is a fundamental challenge in computer security, as attacks are increasing. [31] Focus on malware detection to improve system security. The methodology for this study involves trends to distinguish normal and malicious behaviors. This study is able to accurately predict correct matches and to detect threats. Ransomware is a malware that collects users' data and requires payment to restore access.

Cyber threats have evolved from a severe form of threat affecting systems, resulting in increased economic and data losses globally. The Traditional detection techniques have indicated limitations in identifying and alleviating cyber threats. [32] Propose a machine learning technique that combines a software platform with a hardware platform to improve detection performance. The proposed technique utilizes the neural architecture search method to program the models, sufficiently improving the system. The study indicates that the proposed method enhances detection accuracy and decreases detection delay compared to current methods. [33] present an Ensemble Classifiers technique for Ransomware Detection in Cybersecurity (VBECM-RDCS). This method uses a model for feature selection. When the method is evaluated, it achieves a more accurate result than current methods.

This study presents a method for monitoring malicious threats, modeling changes to the model [34] . Using the extended Berkeley Packet Filter technique, this method enables a reduced performance load while introducing usable, real-time detection trends. This study is based on real-world events. [35] proposed the Self-Modifying Dynamic Pushdown Network (SM-DPN) method for investigating problems in multi-threaded parallel programs. This is a system of Self-Modifying Pushdown networks. A technique is presented to conduct a backward reachability analysis of SM-DPNs. This study introduces a case study and proposes how this method can be applied to evaluate this system affected by malware.

### 3. Methodology

This study untakes an extensive review of past literature in the previous section. This section describes the data collection method, the various techniques used for categorizing malware detection, and the comparisons of these techniques.

### 3.1. Data Collection

This section presents the procedure in acquiring the Android Applications dataset with a description of the dataset. The data for this research were collected from the Kaggle repository, specifically the Drebin dataset, and consist of malicious and benign categories obtained from different applications. The proposed application is designed using Unified Modeling Language (UML). The data consists of attributes extracted from instances, of which are malicious applications, while others are benign.

Table 1 presents characteristics of the dataset distribution and projects the Android malware and benign datasets. Every row has a special identifier that is an API call signature, Intent, Manifest Permission, or Command Signature. This is displayed in Figure 2.

**Table 1.** Dataset Character Distribution

| S/N | CATEGORY | FEATURE |
|---|---|---|
| 1 | Manifest Permissions | 113 |
| 2 | API Call Signatures | 73 |
| 3 | Command Signatures | 6 |
| 4 | Intents | 23 |
| | TOTAL | 215 |

| Name | Type | Description |
|---|---|---|
| RECEIVE_BOOT_COMPLETED | Name | Manifest Permission |
| RESTART_PACKAGES | Type | Manifest Permission |
| LJava.lang.Class.getPackage | Type | API call signature |
| chmod | Type | API call signature |
| LJava.lang.Class.getDeclaredClasses | Name | Intent |
| android.intent.action.ACTION_POWE | Intent | Intent |
| android.intent.action.PACKAGE_ADD | Intent | Intent |
| PathPathLoader | Type | API call signature |
| TelephonyManager.getSimSerialNum | Type | API call signature |
| BLUETOOTH | Type | API call signature |
| READ_CALENDAR | Type | Manifest Permission |
| READ_CALL_LOG | Type | Manifest Permission |
| SUBSCRIBED_FEEDI WRITE | Type | Manifest Permission |
| TelephonyManager.getSimCountryIso | Type | API call signature |
| PackageInstaller | Intent | API call signature |
| sendDataMessage | Type | Manifest Permission |
| sendDataMessage | Intent | Manifest Permission |
| sendDatam | Type | API call signature |
| sendDataAs | Intent | Manifest Permission |
| android.intent.action.PACKAGE_DA | Intent | API call signature |
| sendDataMessage | Intent | Intent |
| ACCESS_NETWORK_STA | Type | Manifest Permission |
| sendDataMessage | Intent | Manifest Permission |

**Figure 2.** Dataset Features

Figure 3 presents the characteristics of the dataset distribution, which is in the form malware or benign). These serve as the inputs to the machine learning model.

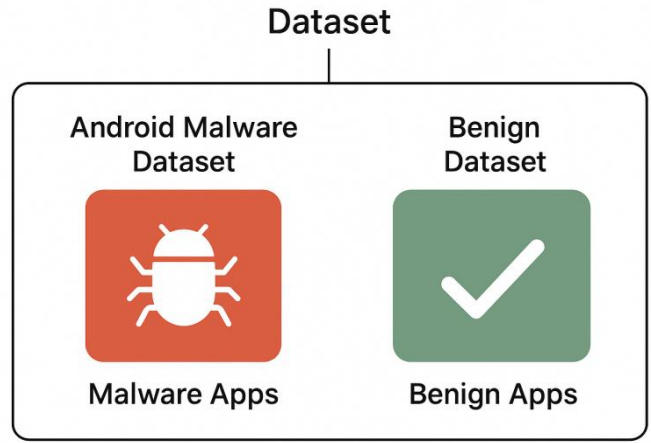

**Figure 3.** Android Malware and Benign Dataset

### 3.2. Techniques for Categorizing Malware Detection

This section describes the various categories of malware analysis techniques, including definitions and research descriptions of the study. The rules and framework for the techniques are introduced. The algorithms for the methods are outlined in the following sections.

#### 3.2.1. Static Analysis

This technique enables analysis to be carried out before program execution [36]. With static analysis, the application can be executed normally, without any breaks. Once a static analysis identifies malware, the issue can be fixed before the application is executed.

The static analysis algorithm is used to resolve issues concerning cyber attacks.

The algorithm consists of four steps [37]:

**Step 1:** A process of unpacking and reorganizing APK files by using a particular packing technique. T. Java-based programs are used for reorganizing the APK files.

The .apk files are reorganized to acquire .dex files.

**Step 2:** The process begins by unpacking the .dex files from APK files.

The process is changed to a directed graph. Every file has its distinct graph design, with every vertex representing an operation. The measurement for each graph is the value for each vertex.

**Step 3**: The sum of all the graphs results in the total graph design.

**Step 4:** The matrices representing the subgraphs are inserted into a static analysis network. A classifier is used to classify the subgraph features.

### 3.2.2. Dynamic Analysis

This technique evaluates the characteristics of a running program. This technique is described by [38]. The dynamic analytical process constitutes an association between objects and attributes. The association is represented as a Boolean table where rows represent the objects and columns the attributes.

The value in the table is true if the object is equal to the attribute and false otherwise.

The objects are tests, and the attributes are the application variables with the test.

The value (T, E) is a notion for every T surrounding E.

T is a series of tests. E is a series of applications.

When a test $t \in T$ is a notion $c$, then

$$t > c \tag{1}$$

If a notion $e$ is equal to $c$, then it is smaller than the notion.

For every notion $e$, there exists a distinct greatest notion that shows and is represented by the notion $c$.

### 3.2.3. Real Time Analysis

The basis of this technique is to have an actual effect for a selected time frame [39]. The real-time system consists of the computer and its environment. The computer associates its environment, depending on the data, with its environment. The real-time system will produce readings at intervals, and the computer will respond by conveying signals, which can be regular or irregular, and must also reply.

The scheduling process for a real-time analysis is given as

$$\tau = \{\tau_1 + \tau_2 + \ldots + \tau_n\} \tag{2}$$

With a set of processors given by

$$\pi = \{\pi_1 + \pi_2 + \ldots + \pi_m\} \tag{3}$$

The release times are given as

$$R = R_1 + R_2 + \ldots + R_k \tag{4}$$

The following represents
$t_i \in \tau$
If ready times are the same, then for all j
$R_j = 0$ for all j
The deadlines for the penalty function are
$D_j$
Completion time is
$C_j$
The priority function is given as
$\Psi_j$
The following parameters are computed.
Flow time as

$$F_j = C_j - R_j \tag{5}$$
$$L_j = C_j - D_j \tag{6}$$

For each time , individual processors are assigned to each task.
If functions are
$\tau_i$ and $\tau_j$   I,j=1,2,...n
The scheduled distance is given as

$$C_{max} - max(c_j) \tag{7}$$

The average flow time is

$$F = \frac{1}{n}\sum_{j=1}^{n} F_j \tag{8}$$

The response time for a system is vital from the user's perspective, as its lowest point yields a lowest mean response time and the mean in-process time of the anticipated response. The scheduling problem is a group of variables described above, together. with the optimal criterion. The benchmarks described are standard, as they require particular techniques for developing the schedules for a particular problem.

### 3.2.4. Hybrid Analysis

It comprises the combination of two methods for malware detection. It has been used extensively to solve problems in malware detection. The study utilizes a hybrid malware classification for more accurate detection. The methods include data set gathering, feature extraction, and modeling methods for malware classification. The algorithm is described by [40]. The steps are given as follows:

**Step 1.** The data for this study comprises a publicly available malware repository and a behavioral analysis dataset. The data contains normal and malicious samples of ransomware, trojans, worms, spyware, and APTs. The following software is utilized for data gathering: VirusTotal, Hybrid Analysis, and Cuckoo Sandbox reports. Preprocessing is applied to data to separate unrelated files.

**Step 2.** The study utilizes Ghidra and IDA Pro for acquiring sequences, control flow graphs, and entropy features for static analysis. The malware is analyzed using Cuckoo Sandbox and Sysmon for dynamic analysis. The extracted features are feature quantities within the opcode sequences in n-gram analysis displayed by graphs, inserted into the embedding algorithm to produce byte-level embeddings for identifying trends in malicious data.

**Step 3:** The Convolutional Neural Network is used to analyze behavior models of malware execution to enhance classification accuracy. The hybrid model is generated for classification using Convolutional Neural Networks and Graph Neural Networks. The study uses malware execution graphs to discover malicious trends.

**Step 4.** The assessment of the methods is done using performance metrics for the classification system. The hybrid model is compared with the conventional machine learning methods to measure the accuracy of the proposed hybrid model.

### 3.3. Comparison of Techniques

This section provides explanations of the comparisons among the four separate categories of malware analysis techniques. The positive and negative attributes for each method are displayed in Table 2. The attributes are examined to determine the characteristics of each method. For this reason, challenges associated with each method are analyzed.

**Table 2.** Comparison of Methods

| Technique | Positive Attributes | Negative Attributes |
|---|---|---|
| Static Analysis | 1 Can securely recognize known malware<br>2 Can fastly recognize known malware<br>3 Applies programming standards<br>4 Easy to maintain | 1 Ineffectual for unknown malware<br>2 Passive to common ambigous techniques<br>3 It takes a lot of time<br>4 Does not support all programming languages<br>4 They geneate false classification values<br>5 Not much qualified human resources |
| Dynamic Analysis | 1 Can recognize unknown malware<br>2 Permits deep insight into results<br>3 outstanding application to research | 1 Tough for ambigous techniques<br>2 Takes too much time for analysis<br>3 Generates excessive classification values |
| Real Time | 1 Fast fending off of malicious threats<br>2 Consistent and potent monitoring of threats<br>3 Powerful for detecting developing threats<br>4 Quicker decision making | 1 Requires much facilities<br>2 Reduced quality of data<br>3 Requires quality in personnel |
| Hybrid | 1 Integrates positives of static and dynamic techniques<br>2 Prevents disadvantages of static and dynamic techniques<br>3 Improved efficiecy<br>4 Lower costs | 1 It requires much facilities<br>2 Sophiscated installation<br>3 Managing the system can be complex<br>4 Expensive maintainance |

### 3.4. Evaluation of Techniques

This section compares the proposed model with other models, evaluating their results with those from these models, using their respective error values. The process of validation of results was discussed by [41]. The computation of errors was described.

$$RMSE = \sqrt{\sum \{e_t\}^2}$$

$$MAPE = \frac{\sum \frac{|e_t|}{x_t}}{n}$$

$et$ are the errors at period $t$, $t$ is the period, $x_t$ is the current value, for n observations.

## 4. Results

The outcomes of this research are in this section. The errors are computed as described in Section 3.4. The output is implemented using the MATLAB programming language (Appendix 1). Tables 3 and 4 focus on the reliability of values for the individual models by calculating their respective error values.

**Table 3.** Comparisons

| Real time | | Dynamic | | Hybrid | | Static | |
|---|---|---|---|---|---|---|---|
| Real | Evaluated | Real | Evaluated | Real | Evaluated | Real | Evaluated |
| 41.72 | 48.93 | 41.33 | 45.78 | 41.11 | 46.78 | 46.65 | 51.78 |
| 45.60 | 50.43 | 45.60 | 48.77 | 45.60 | 49.91 | 43.60 | 51.71 |
| 42.74 | 46.67 | 42.74 | 45.14 | 42.74 | 45.94 | 47.81 | 49.22 |
| 44.75 | 50.95 | 44.75 | 48.32 | 44.75 | 49.01 | 46.55 | 51.57 |
| 41.18 | 88.21 | 41.18 | 45.34 | 41.18 | 46.22 | 43.18 | 48.34 |
| 36.79 | 42.66 | 36.79 | 39.21 | 36.79 | 40.54 | 46.88 | 52.68 |
| 41.68 | 47.79 | 41.68 | 45.11 | 41.68 | 46.32 | 44.78 | 48.90 |
| 42.84 | 46.41 | 42.84 | 44.32 | 42.84 | 44.99 | 44.67 | 48.43 |
| 44.03 | 47.76 | 44.03 | 45.01 | 44.03 | 45.87 | 48.32 | 47.45 |
| 43.78 | 49.46 | 43.78 | 45.32 | 43.78 | 46.58 | 45.66 | 48.44 |
| 44.94 | 48.42 | 44.94 | 45.11 | 44.94 | 46.36 | 46.54 | 49.44 |
| 42.12 | 47.68 | 42.12 | 43.92 | 42.12 | 44.93 | 44.44 | 43.92 |
| 37.91 | 42.43 | 37.91 | 39.88 | 37.91 | 40.03 | 37.91 | 39.88 |
| 42.95 | 47.74 | 42.95 | 44.43 | 42.95 | 45.46 | 42.95 | 44.43 |
| 43.56 | 49.32 | 43.56 | 46.47 | 43.56 | 47.84 | 43.56 | 46.47 |
| 44.34 | 47.46 | 44.34 | 45.13 | 44.34 | 45.79 | 44.35 | 47.54 |
| 44.64 | 48.34 | 44.64 | 45.73 | 44.64 | 46.74 | 42.47 | 49.67 |
| 44.15 | 48.85 | 44.15 | 45.58 | 44.15 | 46.36 | 44.44 | 47.34 |
| 38.90 | 42.85 | 38.90 | 39.41 | 38.90 | 40.58 | 41.43 | 47.41 |
| 37.76 | 41.47 | 37.76 | 38.74 | 37.76 | 39.74 | 39.11 | 41.23 |
| 43.58 | 47.27 | 43.58 | 44.82 | 43.58 | 45.43 | 44.31 | 47.98 |
| 45.15 | 48.11 | 44.15 | 45.15 | 44.15 | 46.85 | 51.23 | 54.57 |

Table 4. Evaluation of Results

|  | Real time | Hybrid | Dynamic | Static |
|---|---|---|---|---|
| **RMSE** | 0.873 | 0.596 | 0.704 | 0.814 |
| **MAPE (%)** | 1.957 | 0.921 | 1.199 | 1.209 |

## 5. Summary and Conclusion

This section undertakes a comprehensive survey of the methods used for detecting malware in communication networks. The characteristics of the individual method are summarised below. The advantages and disadvantages of each method were determined in the review. The process involved in these methods and their characteristics have been explored. This paper presents the major concepts and positive and negative attributes in malware detection.

A static analysis method is introduced in this study for cyber threat detection, which has the potential to accurately identify cyber threats in a communication network. This method integrated a sequence of graph analyses to improve accuracy. The dynamic analysis technique will assist in program evaluation, testing new software, and help programmers in perceiving program behavior. The hybrid technique is a crucial improvement in detecting malware in communication networks.

Real-time processing systems are fast and efficient systems. This technique enables a fast detection of malware in the system. The real-time system comprises the system and a controlled environment. The system is based on its controlled environment, which is dependent on data available about the environment. The real-time system furnishes readings at intervals. There is a time limit for when responses are received.

This study provides an introduction to a survey of observation-based cyber attacks on secure platforms. They present a background of factual research on enhancing malware detection in the communication networks. The results show the Hybrid model to be the best technique. From Table 4, the most accurate technique is the Hybrid model. This is compared with the Real-time, Dynamic, and Static models. The proposed model has the lowest error values. A study of past literature presented in Section 2 confirms that the techniques combined to form the Hybrid model are efficient . The strengths and weaknesses of these techniques are listed. The combination of the other three techniques considered improves the accuracy of the results. This study proposes a model to detect malicious applications in communication networks, as the results from a set are more accurate than those of a single model.

The future work of this study identifies features for an encryption-type ransomware developed for the Android operating system, with a system based on these features. The use of these features will play a great role in alleviating cyber threats. Thus, the importance of this study. Furthermore, this study will employ a comparison of several techniques for malware analysis data, while identifying the most efficient technique.

**Declaration of Competing Interest**

No conflict.

**Authorship Contribution Statement**

**Ozoh Patrick:** Writing, Reviewing
**Omoniyi John:** Methodology, Editing
**Ibitoye Bukola:** Reviewing and Editing

# Appendix 1

# MATLAB Code for Computing *RMSE* and *MAPE*

```matlab
function V=errperf(T,P,M)
%   rmse (root mean squared error)
%   mape (mean absolute percentage error)
%   e (errors)
%   se (squared errors)
%   ape (absolute percentage errors)
%{
Abbreviations:
e: error(s)
M: METRIC
m: mean
P: PREDICTIONS
p: percentage
s: squared
T: TARGETS
V: VALUE(S)
%}
% Transform input
M=lower(M);
%% Compute metric
switch M
    % Errors
    case 'e'
        V=T-P;
    % Squared errors
    case 'se'
        Ve=errperf(T,P,'e');
        V=Ve.^2;
    % Mean squared error
    case 'mse'
        Vse=errperf(T,P,'se');
        V=mean(Vse);
    % Root mean squared error
    case 'rmse'
        Vmse=errperf(T,P,'mse');
        V=sqrt(Vmse);
    % Relative errors
    case 're'
        assert(all(T),'All elements of T must be nonzero.')
        Ve=errperf(T,P,'e');
        V=Ve./T;
    % Percentage errors
    case 'pe'
        Vre=errperf(T,P,'re');
        V=Vre*100;
    % Absolute percentage errors
    case 'ape'
        Vpe=errperf(T,P,'pe');
        V=abs(Vpe);
    % Mean absolute percentage error
    case 'mape'
        Vape=errperf(T,P,'ape');
        V=mean(Vape);
```